\documentclass[aip,graphicx]{revtex4-1}
\usepackage[official]{eurosym}
\usepackage{graphicx}
\usepackage{float}

\usepackage[dvipsnames]{xcolor}

\draft 

\begin{document}


\title{Development of a high-frequency dielectric spectrometer using a portable vector network analyzer} 



\author{Aitor Erkoreka}
\email{aitor.erkorekap@ehu.eus}
\author{Josu Martinez-Perdiguero}
\affiliation{Department of Physics, Faculty of Science and Technology, University of the Basque Country UPV/EHU, Bilbao, Spain}


\date{26 December 2023}

\begin{abstract}
A simple and novel setup for high-frequency dielectric spectroscopy of materials has been developed using a portable vector network analyzer. The measurement principle is based on radio-frequency reflectometry, and both its capabilities and limitations are discussed. The results obtained on a typical liquid crystal prove that the device can provide reliable spectra between $10^7$ Hz and $10^9$ Hz, thus extending the capabilities of conventional impedance analyzers.

\end{abstract}


\maketitle 
\section{Introduction}
Below the frequencies of resonance phenomena due to atomic or molecular vibrations characteristic of optical processes lies the domain of broadband dielectric spectroscopy (BDS). This technique investigates the electrical properties of materials as a function of frequency. For this purpose, the sample is typically sandwiched between two electrodes and an impedance analyzer measures its response to a sinusoidal electric field. The frequency regime of BDS now extends from $10^{-6}$ Hz up to $10^{12}$ Hz, opening up the possibility of observing a wide variety of so-called relaxation processes.\cite{kremer_broadband_2003} Among these are the reorientational motions of molecular dipoles and the translational motions of free charges (ions and electrons). These fluctuations determine the dielectric properties of the material under study, and a great deal of information about molecular dynamics and structure can be extracted from carefully analyzing the measured spectra. Thanks to the considerable advances it has experienced over the last few decades, BDS is nowadays a valuable tool in condensed matter physics and physical chemistry. In fact, even though infrared (IR) and nuclear magnetic resonance (NMR) spectroscopies are more popular among researchers, BDS is a rapidly growing field and is set to match IR and NMR in the next thirty years.\cite{woodward_broadband_2021}

Depending on the frequency range of interest, several different methods can be used to measure the complex dielectric permittivity $\varepsilon(f)=\varepsilon^{\prime}(f)-i\varepsilon^{\prime \prime}(f)$. Below $10^6$ Hz, Fourier correlation and impedance analysis are the most common methods. At higher frequencies ($10^6-10^{10}$ Hz), radio-frequency (RF) reflectometry and network analysis can be employed. Finally, the higher-frequency end of BDS ($10^{10}-10^{12}$ Hz) is dominated by quasi-optical setups\cite{thz_rev_sci_ins} or even Fourier-transform spectroscopy at the far infrared. Most materials physics laboratories have access to impedance analyzers which, under ideal conditions (i.e. good connections, very careful calibrations and appropriate equipment), can perform experiments up to $10^{6}$ Hz. It is worth commenting here that although this equipment is usually advertized to measure up to $10^7$ Hz, it is not so and $10^6$ Hz is, in our experience, more realistic. However, certain relaxation processes in molecular materials like glass-forming liquids, polymers or liquid crystals occur at slightly higher frequencies ($10^{8}-10^{9}$ Hz). On the one hand, certain experimental setups especially designed for high frequencies are rather sophisticated and their implementation is not devoid of difficulties. On the other hand, specialized impedance or vector network analyzers (VNAs) can bridge this gap and provide measurements up to a few GHz. Nonetheless, these instruments are generally very expensive, most notably in the case of high-quality devices for materials science. The fact that the spectra are only stretched one or two decades at such a high price discourages many laboratories from expanding their capabilities.

The NanoVNA project has recently emerged as a low-cost alternative to conventional VNAs.\cite{nanovna.com, nanorfe.com} It is a portable VNA originally intented for measuring antennas, filters, duplexers and amplifiers. Just like regular VNAs, it can characterize the device under test (DUT) by generating an electromagnetic wave of a given frequency and measuring either the reflected or the transmitted signal. Since its original design, several improvements have been implemented and different versions are currently being commercialized by various manufacturers, some of them at a very low price. Moreover, the NanoVNA is an open-source project, and extensive resources and support are available online.\cite{nanovna_original, nanovna_h, nanovna_qt, nanovna_saver} Up to now, there have been a few works that have carried out permittivity measurements with a NanoVNA device, but only in the context of soil analysis.\cite{gonzalez-teruel_measurement_2022, moret-fernandez_testing_2022, soil_mdpi} Furthermore, these studies were performed using open-ended coaxial probes or rod-like probes, which are not suitable for temperature-dependent measurements or if the amount of available material is somehow limited. Additionally, the authors had to carry out a sophisticated calibration procedure for that purpose. In any case, promising results were obtained. In fact, in one of these studies, the authors measured the dielectric spectra of a few reference liquids and compared the results with an Agilent 4395A VNA.\cite{gonzalez-teruel_measurement_2022} The relative error was small in all cases, and they were able to measure up to $900$ MHz. The NanoVNA version they used, however, has already been superseded, so even better results can be obtained in principle. Indeed, the newer versions are now capable of reaching $4.4$ GHz with a higher dynamic range. Finally, it should be mentioned that other authors have also explored the NanoVNA's capabilities for biotechnological applications.\cite{biotechnological, iaccheri_cost-effective_2022}

The aim of this work is to develop a dielectric spectroscopy setup capable of measuring up to $\sim 10^9$ Hz. In this way, researchers working in the field of relaxation phenomena will be able to stretch dielectric spectra a few more decades in frequency with a simple experimental arrangement, which is also very affordable. The general considerations and working principle of the device will be explained in detail. Lastly, the results on a liquid crystal will be presented together with those of the widespread ALPHA-A impedance analyzer from Novocontrol, leading to a complete characterization of the material.

\section{Methods}

\subsection{NanoVNA}
In this work we used the NanoVNA V2. Similarly to other commercially available models, it has a two-port network implemented with SMA connectors and a touch screen. A micro USB port allows for direct computer connection, from which the device can be controlled and measurements and calibrations saved. The nominal measurement frequency-range of this device is $50$ kHz--$4.4$ GHz, although it will only be used in the 10 MHz--1 GHz range due to the lower accuracy outside this interval. After calibration, the system dynamic range is $70$ dB. The device additionally comes with two SMA cables, male/female adapters and calibration standards (open, short and $50$ $\Omega$ standards).

\subsection{Measurement principle}

\begin{figure}[H]
\centering
\includegraphics[width=0.7\textwidth]{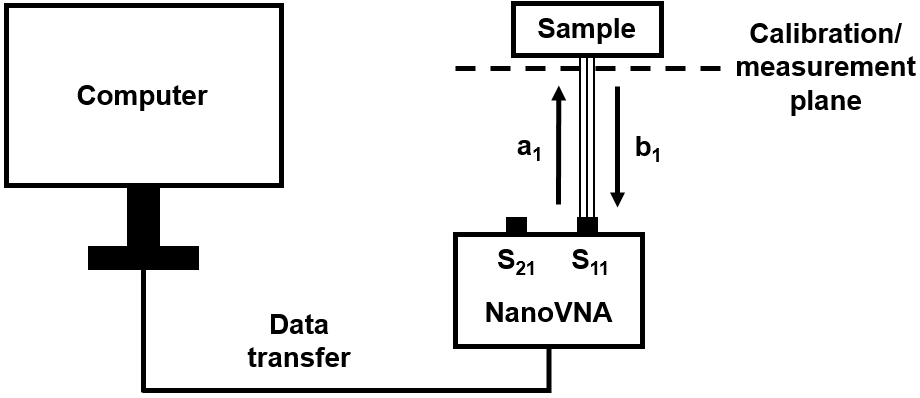}
\caption{\label{fig:scheme} Schematic of the measurement principle.}
\end{figure}

As mentioned previously, the two-port network of the NanoVNA allows measuring on reflection or transmission. In terms of the scattering parameters (S-parameters) typically used in microwave engineering, these are the so-called $S_{11}$ and $S_{21}$ parameters, respectively.\cite{pozar} In the context of BDS, the easiest method is to perform the experiments on reflection. Fig. \ref{fig:scheme} shows the experimental scheme for the dielectric measurements. The sample, like in many BDS experiments, is sandwiched between two electrodes forming a parallel-plate capacitor. The NanoVNA is connected by means of a coaxial line to the sample, and the ratio of the reflected ($b_{1}$) and incident ($a_{1}$) waves is obtained, namely:\cite{pozar}

\begin{equation}
    S_{11}=\frac{b_{1}}{a_{1}}.
\end{equation}

It is important to note that $S_{11}$ is a complex number. Since this quantity is equivalent to the complex reflection coefficient in this case, we will denote it by the more common symbol $\Gamma$. The complex impedance of the sample can then be calculated as\cite{kremer_broadband_2003, pozar}

\begin{equation}
    Z=Z_0 \frac{1+\Gamma}{1-\Gamma},
\end{equation}

\noindent where $Z_0$ is the system characteristic impedance ($50$ $\Omega$ in most cases). The impedance spectrum can be directly saved in the computer from the NanoVNA-QT program.\cite{nanovna_qt} The complex dielectric permittivity of the sample can then be easily obtained from its impedance:\cite{kremer_broadband_2003}

\begin{equation}
    \varepsilon=\frac{d}{i \omega \varepsilon_0 A Z},
\end{equation}

\noindent where $\omega = 2\pi f$, $\varepsilon_0$ is the vacuum permittiviy, $A$ is the area of the capacitor and $d$ the separation between the plates.

A few issues must be considered when using the NanoVNA. Firstly, prior to any measurement, the device has to be calibrated with short, open and load ($50$ $\Omega$) standards. This calibration needs to be done at the so-called measurement plane (see Fig. \ref{fig:scheme}) in order to minimize propagation losses.\cite{pozar} For this purpose, the frequency interval and the number of sweep points have to be specified. Additionally, the number of averages for the measurements can also be specified. These parameters will be saved for the experiments. Secondly, it should be noted that the NanoVNA has some limitations, and that the characteristics of the capacitor formed by the sample should be taken into account. For instance, it would be convenient to estimate the capacitance of the sample so that its associated impedance falls within the range of the NanoVNA's capabilities. The capacitance of the sample will change with frequency, and one must recall that the device was calibrated to $50$ $\Omega$, so the impedance to be measured should, ideally, not be significantly higher or lower than this value. Finally, the system accuracy at the lower and higher frequency ends will always diminish, so it is best to limit the frequency range for reliable results.

\subsection{Experimental setup}

\begin{figure}
\includegraphics[width=0.65\textwidth]{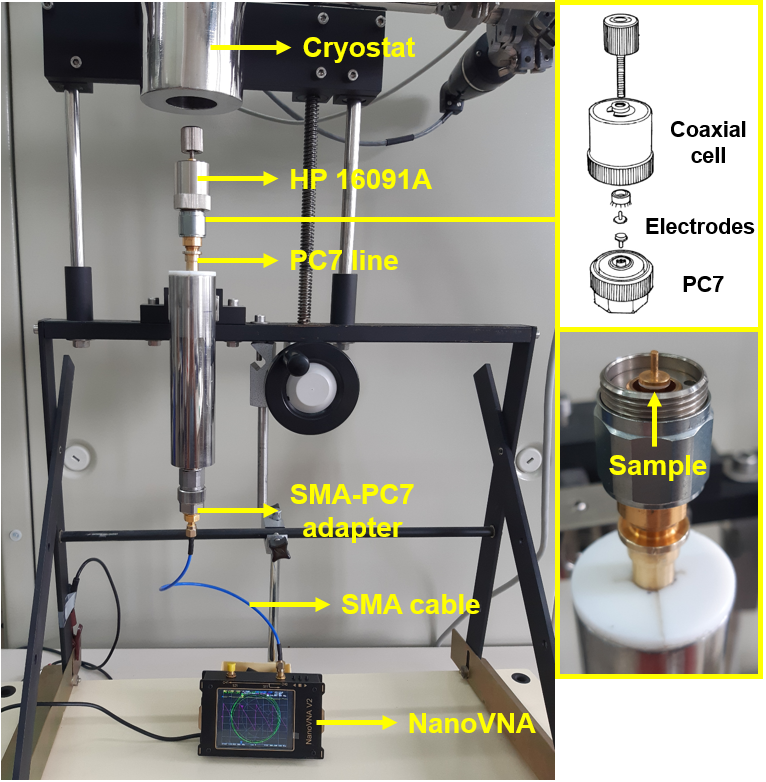}
\caption{\label{fig:setup} Proposed experimental setup for high-frequency dielectric spectroscopy measurements.}
\end{figure}

Constructing an experimental setup based on the measurement principle described above is simple. As already stated, we will use a liquid crystal material as a proof of concept, nonetheless, the setup can be adapted to any kind of sample with very little change. For room temperature measurements, a SMA-PC7 adapter is connected at the end of the SMA cable going from the NanoVNA device to the sample, since the sample is directly connected to a PC7 connector in our fixture. It is important to note that conventional RG58 cables and BNC connectors should not be used to avoid parasitic effects at high frequencies. The sample is sandwiched between two gold-plated brass electrodes, which ensures reliable measurements at high frequencies due to the high conductivity of gold.\cite{parasitic} Since the sample to be measured is a liquid, the electrodes themselves are separated with silica microspheres (20 $\mu$m in diameter from EHC Co. Ltd., Japan). This parallel-plate capacitor is placed at the end of the PC7 connector and the circuit is closed with a modified HP 16091A fixture. In order to keep the connection to the sample vertically straight, a mount with an adjustable clamp can be attached to the PC7 piece. For temperature dependent experiments typical in BDS, this setup must be slightly modified. Instead of placing the sample at the end of the PC7 connector, a PC7 extension line is connected, the top part of which goes inside a cryostat controlled by a Quatro Cryosystem from Novocontrol. The top of the PC7 line is thermally isolated from the bottom by a teflon ring. In this way, the NanoVNA is isolated from the temperature-controlled zone. The sample is placed and the circuit is closed at the end of this line similarly to the room temperature case. The described arrangement is summarized in Fig. \ref{fig:setup}. In any case, as explained earlier, this setup can be greatly simplified for room temperature measurements.

\section{Results and discussion}

\begin{figure}
\centering
\includegraphics[width=0.65\textwidth]{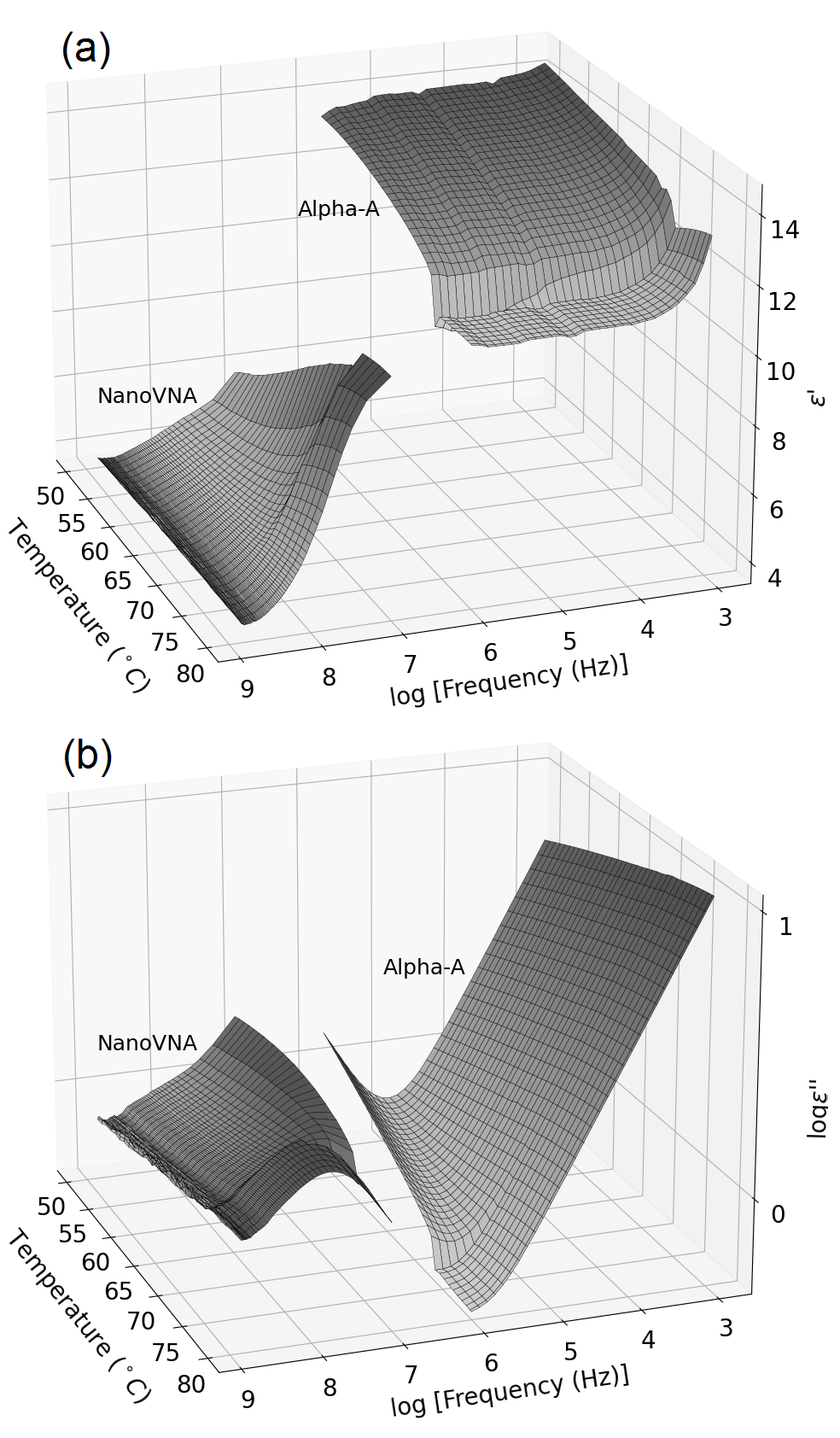}
\caption{\label{fig:3d} Three-dimensional plot of the real (a) and imaginary (b) components of the complex dielectric permittivity as a function of frequency and temperature.}
\end{figure}

In order to test the developed setup, we measured the broadband dielectric spectra of the nematic liquid crystal 7OCB (heptyloxycyanobiphenyl) in the temperature range $50^{\circ}$C--$80^{\circ}$C. The low- and intermediate-frequency measurements ($1$ kHz--$1$ MHz) were carried out using a conventional ALPHA-A impedance analyzer. High-frequency measurements ($10$ MHz--$1$ GHz) were done with the NanoVNA in the proposed configuration. All experiments were done on cooling from $80^{\circ}$C at $0.5^{\circ}$C/min. The results are shown in Fig. \ref{fig:3d}.

\begin{figure}[H]
\centering
\includegraphics[width=0.8\textwidth]{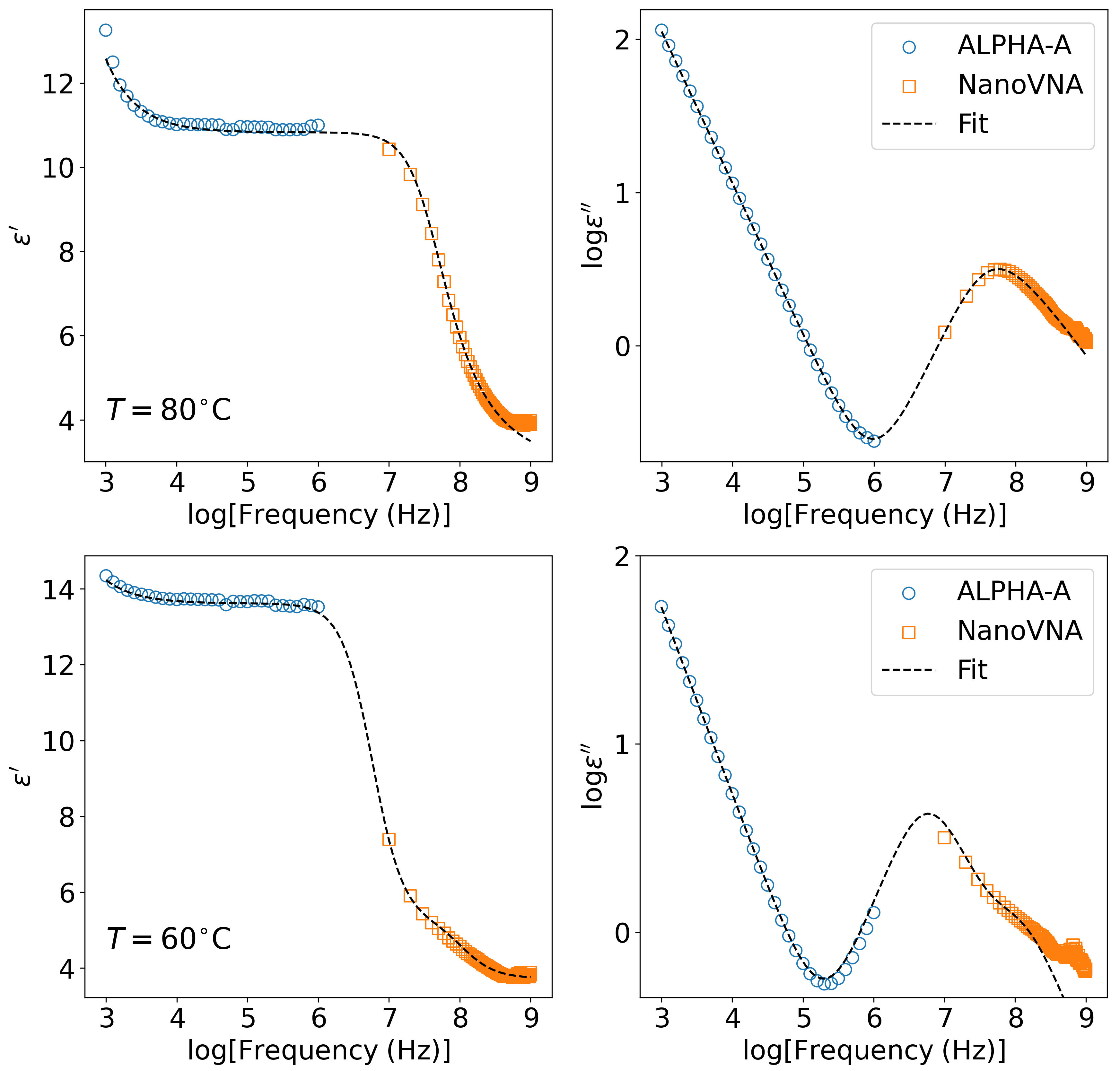}
\caption{\label{fig:spectra} Spectra of the real and imaginary components of the complex dielectric permittivity at $80^{\circ}$C (isotropic phase, top) and $60^{\circ}$C (nematic phase, bottom) measured with the ALPHA-A analyzer and the NanoVNA. The dashed lines correspond to fits to Eq. \ref{HN_eq} with parameters shown in Tables S1 and S2 and RMSE values of 0.019 and 0.021 at $80^{\circ}$C and $60^{\circ}$C respectively.}
\end{figure}

Below $1$ MHz, the spectra are characterized by a plateau in the real component of the permittivity and a pronounced increase in the imaginary component due to ohmic conduction. At the low-frequency end, a slight increase in $\varepsilon'$ is observed, suggesting the onset of the electrode polarization effect. This is a well-known phenomenon that affects the dielectric spectra of a wide variety of samples at low enough frequencies due to the accumulation of ions at the electrodes, and several methods have been developed throughout the years to minimize it.\cite{ep_rev_sci_ins, european, chassagne_compensating_2016} Additionally, a small jump in $\varepsilon'$ is observed at $\sim 74^{\circ}$C, namely, upon the transition from the isotropic to the nematic phase. This indicates that the molecules tend to align somewhat perpendicular to the gold electrodes. At high frequencies ($\sim 10^7$--$10^8$ Hz) another absorption process takes place. Note that the observation of this mode would not have been possible with a standard dielectric spectroscopy setup that can only measure up to $10^6$ Hz. In order to analyze the data more carefully, we plotted the data at two different temperatures (corresponding to the isotropic and nematic phases, respectively) in Fig. \ref{fig:spectra}. These results agree with those already published in the literature.\cite{7ocb} A decrease in the data quality is observed in the higher part of the frequency spectrum, which is expected but which, in our case, is also likely due to the low impedance value (\textit{c.} 2 $\Omega$) of our sample in this highest range. Moreover, although the data are noisier at the high-frequency end, the quality and functional form of the experimental points are adequate, which proves that the setup is suitable to analyze relaxation processes. The data were fitted to the Havriliak-Negami formula with a conductivity term:

\begin{equation}
    \varepsilon (f) = \sum_{k} \frac{\Delta \varepsilon_k}{\left[1+\left(i \frac{f}{f_k}\right)^{\alpha_k} \right]^{\beta_k}} + \varepsilon_{\infty} + \frac{\sigma}{\varepsilon_0(i\,2\pi f)^{\lambda}}\mathrm{,}\label{HN_eq}
\end{equation}

\noindent
where $\Delta\varepsilon_k$, $f_k$, $\alpha_k$ and $\beta_k$ are respectively the dielectric strength, relaxation frequency and broadness exponents of mode $k$, $\varepsilon_{\infty}$ is the high-frequency dielectric permittivity, $\sigma$ is a measure of the conductivity, and $\lambda$ is an exponent between 0 and 1.

At $80^{\circ}$C the data were fitted to a single relaxation process. However, a broadness exponent $\beta = 0.62$ was obtained, suggesting the presence of two independent modes which appear overlapped in frequency.\cite{kremer_broadband_2003} At $60^{\circ}$C, on the contrary, the spectra can be deconvoluted into two relaxation processes: one of them at $\sim 6$ MHz and the other one at $\sim 110$ MHz. These modes, which were almost degenerate in the isotropic phase, correspond to the rotation of individual molecules around their short and long axis, respectively. The fit to the imaginary component at high frequencies is slightly worse due to the already mentioned noise of those data points. Residual plots of both fits in Fig. \ref{fig:spectra} are shown in Fig. S1 in the Supporting Information. The performed deconvolution, as well as the interpretation of the corresponding modes, is in keeping with the literature.\cite{7ocb}

\begin{figure}
\includegraphics[width=0.8\textwidth]{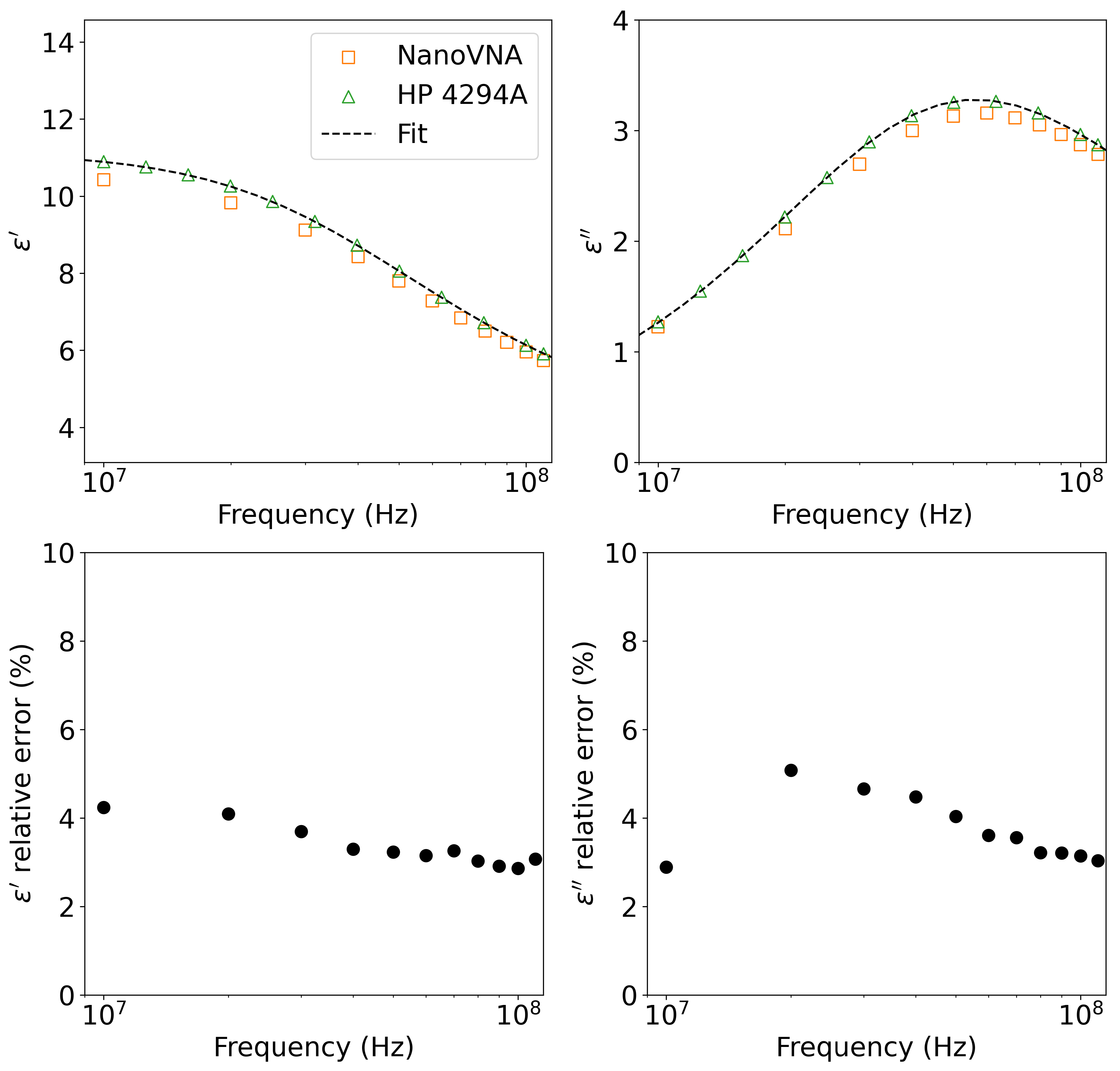}
\caption{\label{fig:comparison} Top: comparison of the measured spectra of the real and imaginary components of the complex dielectric permittivity at $80^{\circ}$C with the NanoVNA and a HP 4294A impedance analyzer. Bottom: calculated relative error with respect to the fitted curve to the data from the HP 4294A impedance analyzer.}
\end{figure}

Although these results agree with measurements already published in the literature, it would be convenient to more carefully assess the accuracy of the experimental data. For that purpose, we performed the same set of experiments with an HP 4294A impedance analyzer, which can measure up to $110$ MHz. In this way, we can at least estimate the quality of the data in the frequency-range $10^7$--$10^8$ Hz. This comparison is shown in the top graphs of Fig. \ref{fig:comparison}, with the data corresponding to the isotropic phase. As it can be seen, the data points lie very close to each other. In order to make this comparison more quantitative, we fitted the data obtained with the HP 4294A analyzer to equation \ref{HN_eq}, taking this as a reference, and calculated the relative error of the NanoVNA data points. The results are plotted at the bottom of Fig. \ref{fig:comparison}. The relative error does not exceed $6\%$ in any case, which is very acceptable since there are sources of error of this order such as the uncertainty in the thickness of the sample. Furthermore, the accuracy of the HP 4294A impedance analyzer in this frequency and impedance range is $\sim 3\%$. Therefore, we conclude once again that the results obtained with the developed setup are good and representative of the material under study.

\section{Conclusions}

A high-frequency dielectric spectrometer for materials analysis based on the reflectometric technique has been developed using a portable VNA. The proposed experimental setup is easy to reproduce and is shown to provide reliable measurements between $10^7$ Hz and $10^9$ Hz. Even though the experimental points near the high-frequency end are slightly noisier owing to the lower dynamic range of the NanoVNA device compared to high-end VNAs/impedance analyzers, the obtained results are adequate. In fact, the quality of the data in the case of a liquid crystal material allows to deconvolute the dielectric spectra into underlying relaxations. We have thus proven the usability and utility of this novel setup for the study of dipolar fluctuations in materials. It is the authors' hope that this work will encourage laboratories working on the dielectric properties of materials to expand their capabilities.

\section{Supplementary material}
The supplementary material includes a document with some information about the fitting quality and parameters as well as files with the raw measurement data.

\section{Acknowledgements}
A.E. and J.M.-P. acknowledge funding from the Basque Government Project IT1458-22. A.E. thanks the Department of Education of the Basque Government for a predoctoral fellowship (grant no. PRE\_2022\_1\_0104).

\section{Author Declaration}
\subsection{Conflict of Interest}
The authors have no conflicts to disclose.

\section{Data Availability}
The data that support the findings of this study are available within the article and its supplementary material.


%
%

%


\bibliography{REFERENCES.bib}

\begin{thebibliography}{20}%
\makeatletter
\providecommand \@ifxundefined [1]{%
 \@ifx{#1\undefined}
}%
\providecommand \@ifnum [1]{%
 \ifnum #1\expandafter \@firstoftwo
 \else \expandafter \@secondoftwo
 \fi
}%
\providecommand \@ifx [1]{%
 \ifx #1\expandafter \@firstoftwo
 \else \expandafter \@secondoftwo
 \fi
}%
\providecommand \natexlab [1]{#1}%
\providecommand \enquote  [1]{``#1''}%
\providecommand \bibnamefont  [1]{#1}%
\providecommand \bibfnamefont [1]{#1}%
\providecommand \citenamefont [1]{#1}%
\providecommand \href@noop [0]{\@secondoftwo}%
\providecommand \href [0]{\begingroup \@sanitize@url \@href}%
\providecommand \@href[1]{\@@startlink{#1}\@@href}%
\providecommand \@@href[1]{\endgroup#1\@@endlink}%
\providecommand \@sanitize@url [0]{\catcode `\\12\catcode `\$12\catcode `\&12\catcode `\#12\catcode `\^12\catcode `\_12\catcode `\%12\relax}%
\providecommand \@@startlink[1]{}%
\providecommand \@@endlink[0]{}%
\providecommand \url  [0]{\begingroup\@sanitize@url \@url }%
\providecommand \@url [1]{\endgroup\@href {#1}{\urlprefix }}%
\providecommand \urlprefix  [0]{URL }%
\providecommand \Eprint [0]{\href }%
\providecommand \doibase [0]{http://dx.doi.org/}%
\providecommand \selectlanguage [0]{\@gobble}%
\providecommand \bibinfo  [0]{\@secondoftwo}%
\providecommand \bibfield  [0]{\@secondoftwo}%
\providecommand \translation [1]{[#1]}%
\providecommand \BibitemOpen [0]{}%
\providecommand \bibitemStop [0]{}%
\providecommand \bibitemNoStop [0]{.\EOS\space}%
\providecommand \EOS [0]{\spacefactor3000\relax}%
\providecommand \BibitemShut  [1]{\csname bibitem#1\endcsname}%
\let\auto@bib@innerbib\@empty
\bibitem [{\citenamefont {Kremer}\ and\ \citenamefont {Schönhals}(2003)}]{kremer_broadband_2003}%
  \BibitemOpen
  \bibinfo {editor} {\bibfnamefont {F.}~\bibnamefont {Kremer}}\ and\ \bibinfo {editor} {\bibfnamefont {A.}~\bibnamefont {Schönhals}},\ eds.,\ \href {\doibase 10.1007/978-3-642-56120-7} {\emph {\bibinfo {title} {Broadband {Dielectric} {Spectroscopy}}}}\ (\bibinfo  {publisher} {Springer Berlin Heidelberg},\ \bibinfo {address} {Berlin, Heidelberg},\ \bibinfo {year} {2003})\BibitemShut {NoStop}%
\bibitem [{\citenamefont {Woodward}(2021)}]{woodward_broadband_2021}%
  \BibitemOpen
  \bibfield  {author} {\bibinfo {author} {\bibfnamefont {W.~H.~H.}\ \bibnamefont {Woodward}},\ }\bibfield  {title} {\enquote {\bibinfo {title} {Broadband {Dielectric} {Spectroscopy}—{A} {Practical} {Guide}},}\ }in\ \href {\doibase 10.1021/bk-2021-1375.ch001} {\emph {\bibinfo {booktitle} {{ACS} {Symposium} {Series}}}},\ Vol.\ \bibinfo {volume} {1375},\ \bibinfo {editor} {edited by\ \bibinfo {editor} {\bibfnamefont {W.~H.~H.}\ \bibnamefont {Woodward}}}\ (\bibinfo  {publisher} {American Chemical Society},\ \bibinfo {address} {Washington, DC},\ \bibinfo {year} {2021})\ pp.\ \bibinfo {pages} {3--59}\BibitemShut {NoStop}%
\bibitem [{\citenamefont {George}, \citenamefont {Charkhesht},\ and\ \citenamefont {Vinh}(2015)}]{thz_rev_sci_ins}%
  \BibitemOpen
  \bibfield  {author} {\bibinfo {author} {\bibfnamefont {D.~K.}\ \bibnamefont {George}}, \bibinfo {author} {\bibfnamefont {A.}~\bibnamefont {Charkhesht}}, \ and\ \bibinfo {author} {\bibfnamefont {N.~Q.}\ \bibnamefont {Vinh}},\ }\bibfield  {title} {\enquote {\bibinfo {title} {{New terahertz dielectric spectroscopy for the study of aqueous solutions}},}\ }\href {\doibase 10.1063/1.4936986} {\bibfield  {journal} {\bibinfo  {journal} {Review of Scientific Instruments}\ }\textbf {\bibinfo {volume} {86}},\ \bibinfo {pages} {123105} (\bibinfo {year} {2015})}\BibitemShut {NoStop}%
\bibitem [{nan({\natexlab{a}})}]{nanovna.com}%
  \BibitemOpen
  \href@noop {} {\enquote {\bibinfo {title} {{N}ano{V}{N}{A} | {V}ery tiny handheld {V}ector {N}etwork {A}nalyzer --- nanovna.com},}\ }\bibinfo {howpublished} {\url{https://nanovna.com/}} ({\natexlab{a}}),\ \bibinfo {note} {[Accessed 11-09-2023]}\BibitemShut {NoStop}%
\bibitem [{nan({\natexlab{b}})}]{nanorfe.com}%
  \BibitemOpen
  \href@noop {} {\enquote {\bibinfo {title} {{N}ano{V}{N}{A} | {N}ano{R}{F}{E} --- nanorfe.com},}\ }\bibinfo {howpublished} {\url{https://nanorfe.com/}} ({\natexlab{b}}),\ \bibinfo {note} {[Accessed 11-09-2023]}\BibitemShut {NoStop}%
\bibitem [{nan({\natexlab{c}})}]{nanovna_original}%
  \BibitemOpen
  \href@noop {} {\enquote {\bibinfo {title} {{G}it{H}ub - ttrftech/{N}ano{V}{N}{A}: {V}ery {T}iny {P}almtop {V}ector {N}etwork {A}nalyzer --- github.com},}\ }\bibinfo {howpublished} {\url{https://github.com/ttrftech/NanoVNA}} ({\natexlab{c}}),\ \bibinfo {note} {[Accessed 11-09-2023]}\BibitemShut {NoStop}%
\bibitem [{nan({\natexlab{d}})}]{nanovna_h}%
  \BibitemOpen
  \href@noop {} {\enquote {\bibinfo {title} {{G}it{H}ub - hugen79/{N}ano{V}{N}{A}-{H}: {N}ano{V}{N}{A}-{H} based on edy555 design, provides effective measurements up to 1.5{G}{H}z. --- github.com},}\ }\bibinfo {howpublished} {\url{https://github.com/hugen79/NanoVNA-H}} ({\natexlab{d}}),\ \bibinfo {note} {[Accessed 11-09-2023]}\BibitemShut {NoStop}%
\bibitem [{nan({\natexlab{e}})}]{nanovna_qt}%
  \BibitemOpen
  \href@noop {} {\enquote {\bibinfo {title} {{G}it{H}ub - nanovna-v2/{N}ano{V}{N}{A}-{Q}{T}: {P}{C} {G}{U}{I} software for {N}ano{V}{N}{A} {V}2 --- github.com},}\ }\bibinfo {howpublished} {\url{https://github.com/nanovna-v2/NanoVNA-QT}} ({\natexlab{e}}),\ \bibinfo {note} {[Accessed 11-09-2023]}\BibitemShut {NoStop}%
\bibitem [{nan({\natexlab{f}})}]{nanovna_saver}%
  \BibitemOpen
  \href@noop {} {\enquote {\bibinfo {title} {{G}it{H}ub - {N}ano{V}{N}{A}-{S}aver/nanovna-saver: {A} tool for reading, displaying and saving data from the {N}ano{V}{N}{A} --- github.com},}\ }\bibinfo {howpublished} {\url{https://github.com/NanoVNA-Saver/nanovna-saver}} ({\natexlab{f}}),\ \bibinfo {note} {[Accessed 11-09-2023]}\BibitemShut {NoStop}%
\bibitem [{\citenamefont {González-Teruel}\ \emph {et~al.}(2022)\citenamefont {González-Teruel}, \citenamefont {Jones}, \citenamefont {Robinson}, \citenamefont {Giménez-Gallego}, \citenamefont {Zornoza},\ and\ \citenamefont {Torres-Sánchez}}]{gonzalez-teruel_measurement_2022}%
  \BibitemOpen
  \bibfield  {author} {\bibinfo {author} {\bibfnamefont {J.~D.}\ \bibnamefont {González-Teruel}}, \bibinfo {author} {\bibfnamefont {S.~B.}\ \bibnamefont {Jones}}, \bibinfo {author} {\bibfnamefont {D.~A.}\ \bibnamefont {Robinson}}, \bibinfo {author} {\bibfnamefont {J.}~\bibnamefont {Giménez-Gallego}}, \bibinfo {author} {\bibfnamefont {R.}~\bibnamefont {Zornoza}}, \ and\ \bibinfo {author} {\bibfnamefont {R.}~\bibnamefont {Torres-Sánchez}},\ }\bibfield  {title} {\enquote {\bibinfo {title} {Measurement of the broadband complex permittivity of soils in the frequency domain with a low-cost {Vector} {Network} {Analyzer} and an {Open}-{Ended} coaxial probe},}\ }\href {\doibase 10.1016/j.compag.2022.106847} {\bibfield  {journal} {\bibinfo  {journal} {Computers and Electronics in Agriculture}\ }\textbf {\bibinfo {volume} {195}},\ \bibinfo {pages} {106847} (\bibinfo {year} {2022})}\BibitemShut {NoStop}%
\bibitem [{\citenamefont {Moret-Fernández}\ \emph {et~al.}(2022)\citenamefont {Moret-Fernández}, \citenamefont {Lera}, \citenamefont {Latorre}, \citenamefont {Tormo},\ and\ \citenamefont {Revilla}}]{moret-fernandez_testing_2022}%
  \BibitemOpen
  \bibfield  {author} {\bibinfo {author} {\bibfnamefont {D.}~\bibnamefont {Moret-Fernández}}, \bibinfo {author} {\bibfnamefont {F.}~\bibnamefont {Lera}}, \bibinfo {author} {\bibfnamefont {B.}~\bibnamefont {Latorre}}, \bibinfo {author} {\bibfnamefont {J.}~\bibnamefont {Tormo}}, \ and\ \bibinfo {author} {\bibfnamefont {J.}~\bibnamefont {Revilla}},\ }\bibfield  {title} {\enquote {\bibinfo {title} {Testing of a commercial vector network analyzer as low-cost {TDR} device to measure soil moisture and electrical conductivity},}\ }\href {\doibase 10.1016/j.catena.2022.106540} {\bibfield  {journal} {\bibinfo  {journal} {CATENA}\ }\textbf {\bibinfo {volume} {218}},\ \bibinfo {pages} {106540} (\bibinfo {year} {2022})}\BibitemShut {NoStop}%
\bibitem [{\citenamefont {Cataldo}\ \emph {et~al.}(2022)\citenamefont {Cataldo}, \citenamefont {Farhat}, \citenamefont {Farrugia}, \citenamefont {Persico},\ and\ \citenamefont {Schiavoni}}]{soil_mdpi}%
  \BibitemOpen
  \bibfield  {author} {\bibinfo {author} {\bibfnamefont {A.}~\bibnamefont {Cataldo}}, \bibinfo {author} {\bibfnamefont {I.}~\bibnamefont {Farhat}}, \bibinfo {author} {\bibfnamefont {L.}~\bibnamefont {Farrugia}}, \bibinfo {author} {\bibfnamefont {R.}~\bibnamefont {Persico}}, \ and\ \bibinfo {author} {\bibfnamefont {R.}~\bibnamefont {Schiavoni}},\ }\bibfield  {title} {\enquote {\bibinfo {title} {A method for extracting {Debye} parameters as a tool for monitoring watered and contaminated soils},}\ }\href {\doibase 10.3390/s22207805} {\bibfield  {journal} {\bibinfo  {journal} {Sensors}\ }\textbf {\bibinfo {volume} {22}} (\bibinfo {year} {2022}),\ 10.3390/s22207805}\BibitemShut {NoStop}%
\bibitem [{\citenamefont {Zaikou}, \citenamefont {Nacke},\ and\ \citenamefont {Pliquett}(2021)}]{biotechnological}%
  \BibitemOpen
  \bibfield  {author} {\bibinfo {author} {\bibfnamefont {Y.}~\bibnamefont {Zaikou}}, \bibinfo {author} {\bibfnamefont {T.}~\bibnamefont {Nacke}}, \ and\ \bibinfo {author} {\bibfnamefont {U.}~\bibnamefont {Pliquett}},\ }\bibfield  {title} {\enquote {\bibinfo {title} {High frequency impedance spectroscopy for biotechnological applications},}\ }in\ \href {\doibase 10.1109/IWIS54661.2021.9711786} {\emph {\bibinfo {booktitle} {2021 International Workshop on Impedance Spectroscopy (IWIS)}}}\ (\bibinfo {year} {2021})\ pp.\ \bibinfo {pages} {70--75}\BibitemShut {NoStop}%
\bibitem [{\citenamefont {Iaccheri}, \citenamefont {Varani},\ and\ \citenamefont {Ragni}(2022)}]{iaccheri_cost-effective_2022}%
  \BibitemOpen
  \bibfield  {author} {\bibinfo {author} {\bibfnamefont {E.}~\bibnamefont {Iaccheri}}, \bibinfo {author} {\bibfnamefont {M.}~\bibnamefont {Varani}}, \ and\ \bibinfo {author} {\bibfnamefont {L.}~\bibnamefont {Ragni}},\ }\bibfield  {title} {\enquote {\bibinfo {title} {Cost-{Effective} {Open}-{Ended} {Coaxial} {Technique} for {Liquid} {Food} {Characterization} by {Using} the {Reflection} {Method} for {Industrial} {Applications}},}\ }\href {\doibase 10.3390/s22145277} {\bibfield  {journal} {\bibinfo  {journal} {Sensors}\ }\textbf {\bibinfo {volume} {22}},\ \bibinfo {pages} {5277} (\bibinfo {year} {2022})}\BibitemShut {NoStop}%
\bibitem [{\citenamefont {Pozar}(2005)}]{pozar}%
  \BibitemOpen
  \bibfield  {author} {\bibinfo {author} {\bibfnamefont {D.~M.}\ \bibnamefont {Pozar}},\ }\href {https://cds.cern.ch/record/882338} {\emph {\bibinfo {title} {{Microwave engineering; 3rd ed.}}}}\ (\bibinfo  {publisher} {Wiley},\ \bibinfo {address} {Hoboken, NJ},\ \bibinfo {year} {2005})\BibitemShut {NoStop}%
\bibitem [{\citenamefont {Perkowski}(2021)}]{parasitic}%
  \BibitemOpen
  \bibfield  {author} {\bibinfo {author} {\bibfnamefont {P.}~\bibnamefont {Perkowski}},\ }\bibfield  {title} {\enquote {\bibinfo {title} {The parasitic effects in high-frequency dielectric spectroscopy of liquid crystals – the review},}\ }\href {\doibase 10.1080/02678292.2020.1852619} {\bibfield  {journal} {\bibinfo  {journal} {Liquid Crystals}\ }\textbf {\bibinfo {volume} {48}},\ \bibinfo {pages} {767--793} (\bibinfo {year} {2021})}\BibitemShut {NoStop}%
\bibitem [{\citenamefont {Pizzitutti}\ and\ \citenamefont {Bruni}(2001)}]{ep_rev_sci_ins}%
  \BibitemOpen
  \bibfield  {author} {\bibinfo {author} {\bibfnamefont {F.}~\bibnamefont {Pizzitutti}}\ and\ \bibinfo {author} {\bibfnamefont {F.}~\bibnamefont {Bruni}},\ }\bibfield  {title} {\enquote {\bibinfo {title} {{Electrode and interfacial polarization in broadband dielectric spectroscopy measurements}},}\ }\href {\doibase 10.1063/1.1364663} {\bibfield  {journal} {\bibinfo  {journal} {Review of Scientific Instruments}\ }\textbf {\bibinfo {volume} {72}},\ \bibinfo {pages} {2502--2504} (\bibinfo {year} {2001})}\BibitemShut {NoStop}%
\bibitem [{\citenamefont {Emmert}\ \emph {et~al.}(2011)\citenamefont {Emmert}, \citenamefont {Wolf}, \citenamefont {Gulich}, \citenamefont {Krohns}, \citenamefont {Kastner}, \citenamefont {Lunkenheimer},\ and\ \citenamefont {Loidl}}]{european}%
  \BibitemOpen
  \bibfield  {author} {\bibinfo {author} {\bibfnamefont {S.}~\bibnamefont {Emmert}}, \bibinfo {author} {\bibfnamefont {M.}~\bibnamefont {Wolf}}, \bibinfo {author} {\bibfnamefont {R.}~\bibnamefont {Gulich}}, \bibinfo {author} {\bibfnamefont {S.}~\bibnamefont {Krohns}}, \bibinfo {author} {\bibfnamefont {S.}~\bibnamefont {Kastner}}, \bibinfo {author} {\bibfnamefont {P.}~\bibnamefont {Lunkenheimer}}, \ and\ \bibinfo {author} {\bibfnamefont {A.}~\bibnamefont {Loidl}},\ }\bibfield  {title} {\enquote {\bibinfo {title} {Electrode polarization effects in broadband dielectric spectroscopy},}\ }\href {\doibase 10.1140/epjb/e2011-20439-8} {\bibfield  {journal} {\bibinfo  {journal} {The European Physical Journal B}\ }\textbf {\bibinfo {volume} {83}},\ \bibinfo {pages} {157--165} (\bibinfo {year} {2011})}\BibitemShut {NoStop}%
\bibitem [{\citenamefont {Chassagne}\ \emph {et~al.}(2016)\citenamefont {Chassagne}, \citenamefont {Dubois}, \citenamefont {Jiménez}, \citenamefont {van~der Ploeg},\ and\ \citenamefont {van Turnhout}}]{chassagne_compensating_2016}%
  \BibitemOpen
  \bibfield  {author} {\bibinfo {author} {\bibfnamefont {C.}~\bibnamefont {Chassagne}}, \bibinfo {author} {\bibfnamefont {E.}~\bibnamefont {Dubois}}, \bibinfo {author} {\bibfnamefont {M.~L.}\ \bibnamefont {Jiménez}}, \bibinfo {author} {\bibfnamefont {J.~P.~M.}\ \bibnamefont {van~der Ploeg}}, \ and\ \bibinfo {author} {\bibfnamefont {J.}~\bibnamefont {van Turnhout}},\ }\bibfield  {title} {\enquote {\bibinfo {title} {Compensating for {Electrode} {Polarization} in {Dielectric} {Spectroscopy} {Studies} of {Colloidal} {Suspensions}: {Theoretical} {Assessment} of {Existing} {Methods}},}\ }\href@noop {} {\bibfield  {journal} {\bibinfo  {journal} {Frontiers in Chemistry}\ }\textbf {\bibinfo {volume} {4}} (\bibinfo {year} {2016})}\BibitemShut {NoStop}%
\bibitem [{\citenamefont {Diez}\ \emph {et~al.}(2006)\citenamefont {Diez}, \citenamefont {Pérez-Jubindo}, \citenamefont {{de la Fuente}}, \citenamefont {López}, \citenamefont {Salud},\ and\ \citenamefont {Tamarit}}]{7ocb}%
  \BibitemOpen
  \bibfield  {author} {\bibinfo {author} {\bibfnamefont {S.}~\bibnamefont {Diez}}, \bibinfo {author} {\bibfnamefont {M.}~\bibnamefont {Pérez-Jubindo}}, \bibinfo {author} {\bibfnamefont {M.}~\bibnamefont {{de la Fuente}}}, \bibinfo {author} {\bibfnamefont {D.}~\bibnamefont {López}}, \bibinfo {author} {\bibfnamefont {J.}~\bibnamefont {Salud}}, \ and\ \bibinfo {author} {\bibfnamefont {J.}~\bibnamefont {Tamarit}},\ }\bibfield  {title} {\enquote {\bibinfo {title} {On the influence of cylindrical sub-micrometer confinement on heptyloxycyanobiphenyl {(7OCB)}. {A} dynamic dielectric study},}\ }\href {\doibase https://doi.org/10.1016/j.cplett.2006.03.072} {\bibfield  {journal} {\bibinfo  {journal} {Chemical Physics Letters}\ }\textbf {\bibinfo {volume} {423}},\ \bibinfo {pages} {463--469} (\bibinfo {year} {2006})}\BibitemShut {NoStop}%
\end{thebibliography}%

\end{document}